\documentclass[9pt,twocolumn]{article}

\usepackage[nocompress]{cite}
\usepackage{graphicx}
\usepackage{amsmath}
\usepackage{algorithmic}
\usepackage[caption=false,font=normalsize,labelfont=sf,textfont=sf]{subfig}
\usepackage{url}
\usepackage{hyperref}
\usepackage{breakurl}
\usepackage{listings}
\usepackage{flushend}
\usepackage[utf8]{inputenc}
\usepackage[T1]{fontenc}
\usepackage[english]{babel}
\usepackage{multirow}
\usepackage{authblk}

\newcommand{\circled}[1]{\raisebox{.4pt}{\textcircled{\raisebox{-1.0pt}{#1}}}}

\begin{document}

\title{Fault Attacks on Encrypted General Purpose Compute Platforms}

\author[3]{Robert Buhren}
\author[1,2]{Shay Gueron}
\author[3]{Jan Nordholz}
\author[4]{Jean-Pierre Seifert}
\author[3]{Julian Vetter}
\affil[1]{{\tt shay@math.haifa.ac.il}, University of Haifa, Israel}
\affil[2]{Intel Corporation, Intel Development Center, Israel}
\affil[3]{{\tt \{robert, jnordholz, julian\}@sec.t-labs.tu-berlin.de}, TU Berlin, Germany}
\affil[4]{{\tt jean-pierre.seifert@telekom.de}, TU Berlin, Germany}

\maketitle
\begin{abstract}
Adversaries with physical access to a target platform can perform cold boot or DMA attacks to extract sensitive data from the RAM.
In response, several main-memory encryption schemes have been proposed to prevent such attacks.
Also hardware vendors have acknowledged the threat and already announced respective hardware extensions.
Intel's SGX and AMD's SME will provide means to encrypt parts of the RAM to protect security-relevant assets that reside there.

Encrypting the RAM will protect the user's content against passive eavesdropping.
However, the level of protection it provides in scenarios that involve an adversary who is not only able to read from RAM but can also change content in RAM is less clear.
Obviously, encryption offers some protection against such an ``active'' adversary:
from the ciphertext the adversary cannot see what value is changed in the plaintext, nor predict the system behaviour based on the changes.
But is this enough to prevent an active adversary from performing malicious tasks?

This paper addresses the open research question whether encryption alone is a dependable protection mechanism in practice when considering an active adversary.
To this end, we first build a software based memory encryption solution on a desktop system which mimics AMD's SME.
Subsequently, we demonstrate a proof-of-concept fault attack on this system, by which we are able to extract the private RSA key of a GnuPG user.
Our work suggests that transparent memory encryption is not enough to prevent active attacks.
\end{abstract}

\section{Introduction}
\label{sec:introduction}
Adversaries use cold boot attacks~\cite{frost2013, halderman2009, gruhn2013}, bus monitoring~\cite{gogniat2008}, and DMA attacks~\cite{becher2005, boileau2006} to steal data from main memory.
Such attacks can be used for capturing the information that happen to populate the RAM at the time of the attack, e.g., keys and other sensitive information.
Opportunities to launch physical attacks present themselves on portable devices such as laptops and smartphones, which can get easily lost or stolen, but also on desktop systems in untrusted environments (e.g. corporate workstations or university computers).
However, these attacks are ``static'' in nature: only a single RAM snapshot, taken at an arbitrary point in time, is available to the adversary.
As a response, transparent encryption of the memory, while the platform is operating (with a secret key that is not stored in RAM), leaves the adversary with only one snapshot of ciphertext, therefore providing perfect mitigation.
Hardware vendors acknowledged this threat as well, with AMD announcing new processor extensions to mitigate such threats.
AMD's SME (Secure Memory Encryption)~\cite{sme2016} provides ways to encrypt parts of the RAM and leave the adversary only with ciphertext.

Although the above mechanisms protect the privacy of the user's data, it is less clear whether such transparent encryption is enough against an active adversary.
Looking at previous work (e.g., ~\cite{funderbolt2013, becher2005, boileau2006}) reveals that a number of different hardware interfaces, e.g., Thunderbolt, Firewire, PCIe, PCMCIA and super-speed USB ports, are readily available and can be used for attack purposes.
In such a scenario it is important to realize that the tools (e.g., DMA devices) that an adversary can use for reading the RAM content can also be used for overwriting the RAM.
With these capabilities, reading cleartext memory and then overwriting it, an adversary can modify any known value on the RAM to any desired value.
The security consequences are obvious.
Such threats can be mitigated by adding integrity protection to the RAM.
Unfortunately, adding integrity to the RAM is complex~\cite{owen2013, elbaz2009}, because it involves generating additional memory transactions on top of the regular data transfer, thus changing the amount of data per read and write.
It also requires dedicated storage on the RAM, for integrity tags, consuming an overhead of $>$20\%.

Thus, the above discussion leads to the following question, which is the focus of this paper.
Memory encryption is required for privacy, and authentication is required for integrity.
But, can transparent memory encryption protect the system from active attacks and obviate the need for expensive authentication?
The logic behind this hope is clear.
The presence of encryption limits the active adversary in a fundamental way, by blinding him in two ways.
He does not know what data is being changed on the encrypted RAM, and -- more importantly -- he has no control on the resulting value when the modified RAM is decrypted.
The security properties in our scenario cannot be proven, so the discussion is reduced to the question of the practicality of two-way blinded attacks. 

We address the problem here and show that the protection offered by transparent encryption against active adversaries is not guaranteed, even from a practical viewpoint.
The main contributions of our paper are the following:
\begin{itemize}
\setlength{\itemsep}{-2pt}
\item{
As platforms with SME are not yet available, we designed a memory encryption scheme that replicates the functionality SME implements in hardware. Our software SME prototype is embedded into the Linux kernel.}
\item{
We built a proof-of-concept fault attack\footnote{We do not expose a real vulnerability in GnuPG.
For our attack, we use GnuPG version 1.4.19, and invoke it with a non-default (though possible) command line that skips the signature checking.
But note that other crypto systems are also threatened by our method, cf.~\cite{joye2012}.} on GnuPG~\cite{gnupg2016}.
Our attack targets the RSA signing procedure of GnuPG.
With our attack we are able to reveal the private RSA key of a GnuPG user.}
\item{
We employed a mechanism based on LLC (Last Level Cache) probing to determine the exact time where the victim process executes code from a specific memory location.
We combine this information with a kernel page allocator prediction mechanism to inject a fault into the victim application's encrypted data in order to cause a predictable effect.}
\item{
We discuss the challenges that an adversary needs to overcome in order to extend our proof-of-concept attack to a real attack.
We also provide several mitigation techniques to prevent the attack.}
\end{itemize}

The paper is organized as follows.
Section~\ref{sec:pre_not} establishes some notation and preliminaries, and
Section~\ref{sec:background} provides the technical background.
We define our attack model in Section~\ref{sec:attack_model}.
We describe our software-based memory encryption prototype and the attack on GnuPG in Section~\ref{sec:soft_seal} and Section~\ref{sec:attack_gnupg}, respectively.
We discuss the challenges of a real-world attack in Section~\ref{sec:real_world_issues}, and we continue by presenting our results in Section~\ref{sec:attack_results}.
In Section~\ref{sec:counter_related_work} we discuss mitigation techniques and related work.
We conclude our work in Section~\ref{sec:conclusion}.

\section{The Boneh-DeMillo-Lipton fault attack on RSA-CRT}
\label{sec:pre_not}
Consider an RSA cryptosystem with a public key $n=pq$ that is the product of two (randomly chosen) secret distinct primes of the same bit-length $k$, $|p|=|q|=k$. 
The public and the private exponents are denoted by $e$ and $d$, respectively, where 
$e\cdot d \equiv 1 \mod (p-1) \cdot (q-1)$.
A standard choice is $e=2^{16}+1$ ($|e|=17$) and $k=1024$, i.e., $|n|=|d|=2048$. 

A signature ($s$) on a message ($m$) is calculated by $s \equiv m^{d} \mod n$ (in applications, $m$ is constructed by concatenating a hash digest with some padding pattern, and interpreting the resulting bit sequence as a $2k$-bit integer).
Verification of a signature $s$ on the message $m$ is carried out by checking that $s^{e} \equiv m \mod n$.
Since $e$ is short, the time it takes to verify a signature is much shorter than the time it takes to create it. 

Almost all efficient implementations use the Chinese Remainder Theorem (CRT).
Secret integers are derived (once) from the private key as follows. 
\begin{eqnarray}
\label{eq:crt_precomp}
d_{p} \equiv d \mod p-1, ~~ d_{q} \equiv d \mod q-1 \\ 
q_{inv} \equiv q^{-1} \mod p
\end{eqnarray}
With these values, $s \equiv m^d \mod n$ can be computed by CRT and Garner's recombination by the following steps. 
\begin{eqnarray}
s_{p} \equiv m^{d_{p}} \mod p, ~~ s_{q} \equiv m^{d_{q}} \mod q \label{eq:crt_precomp1} \\
h_{q} \equiv q_{inv} \cdot (s_{p} - s_{q}) \mod p \label{eq:crt_precomp2} \\
s = s_{p} + h_{q} \cdot q \label{eq:crt_precomp3}
\end{eqnarray}
In Equation \ref{eq:crt_precomp1}, $m$ can be replaced by $m \mod p$ and $m \mod q$, and it follows that the computation of $s$ is dominated by two modular exponentiations of inputs with bit-lengths $k$. This is 4 times faster than the direct computation $s \equiv m^d \mod n$ that involves one modular exponentiation of inputs with bit-lengths $2k$. Due to this performance gain, the use of CRT for RSA signature (and decryption) is the preferred choice.

In this paper, we use the Boneh-DeMillo-Lipton fault attack~\cite{boneh1997}, which can be applied to a device that computes RSA signatures using the CRT.
The attack is based on obtaining two signatures of the same message $m$.
The first one is correct, and denoted by $s$.
The second one is faulty, and is obtained by injecting some corruption (to the computing apparatus), that is {\em timed appropriately} so that the value of $s_q$ is computed correctly, but $s_p$ is corrupted to $s'_{p}$. The recombination (\ref{eq:crt_precomp3}) yields the faulty signature $s'$.
It satisfies (with very high likelihood) $q = gcd(s' - s, n)$, thus leading to factorizing $n$ and hence to discovering the secret exponent $d$.

\section{Technical Background}
\label{sec:background}
This section describes some background that is crucial for understanding the rest of the paper without requiring the reader to be a priori familiar with these details.

\subsection{Memory Management}
\label{sec:mm}
Most general purpose compute platforms provide a hardware MMU (Memory Management Unit), whereby the OS is able to implement the virtual memory abstraction.
The OS uses virtual memory to organize its address space and relies on the MMU's address translation mechanism for implementing the necessary isolation between unprivileged applications and the OS.
These protection measures prevent bugs or malicious activity in one application from impacting other applications or the OS kernel itself.
Each process uses a separate virtual address space that references only the memory allocated to that process.
From the viewpoint of an application that uses virtual memory, it looks like it is running on a separate computer and has its own RAM.
From a system's perspective, the address translation is a layer of indirection between virtual addresses and physical addresses.

\subsection{Cache architecture}
\label{sec:cache}
Modern x86 processors have multiple levels of caches, which are structured in a hierarchical manner.
Each core on the system has its own dedicated L1 and L2 cache.
All caches except for the L1 cache are unified, storing data and instructions.
The L3 cache or LLC (Last Level Cache) is usually shared among all CPU cores on the chip.
However Intel splits the LLC into several parts where each CPU core has a local part of the LLC and can access remote parts with an increased latency.

The cache is divided memory chunks of equal size, the so-called cache lines.
On a x86-64 system, the typical size of a cache line is 64~Bytes.
The x86-64 caches operate in a set-associative mode.
All available slots are grouped into sets of a specific size.
This number varies, depending on the processor and the size of the cache.
Each memory chunk can be stored in all slots of one particular set.

The addressing of the cache is determined by various bits of the physical address.
The lowest bits of each address denote the offset inside the cache line.
The intermediate bits determine the set.
The remaining high bits form the address tag, that has to be stored with each cache line for the later lookup.
Additionally, when looking at the Sandy Bridge LLC, the high address bits are
also taken into consideration for the calculation of the cache slice~\cite{hund2013}.

It can be observed that, when looking at the set-associativity, memory addresses with identical index bits compete on the available slots of one set.
Hence, memory accesses may evict and replace other memory content from the caches.

\section{Attack model}
\label{sec:attack_model}
In this paper we consider general purpose compute platforms (e.g. desktop computers or laptops) which are located in an untrusted environment.
For such a scenario AMD's memory encryption provides an appealing option.
SME is a general purpose mechanism to encrypt the RAM, which works on desktops, laptops and workstation systems.

\begin{figure}
  \centering
  \includegraphics[width=.9\linewidth]{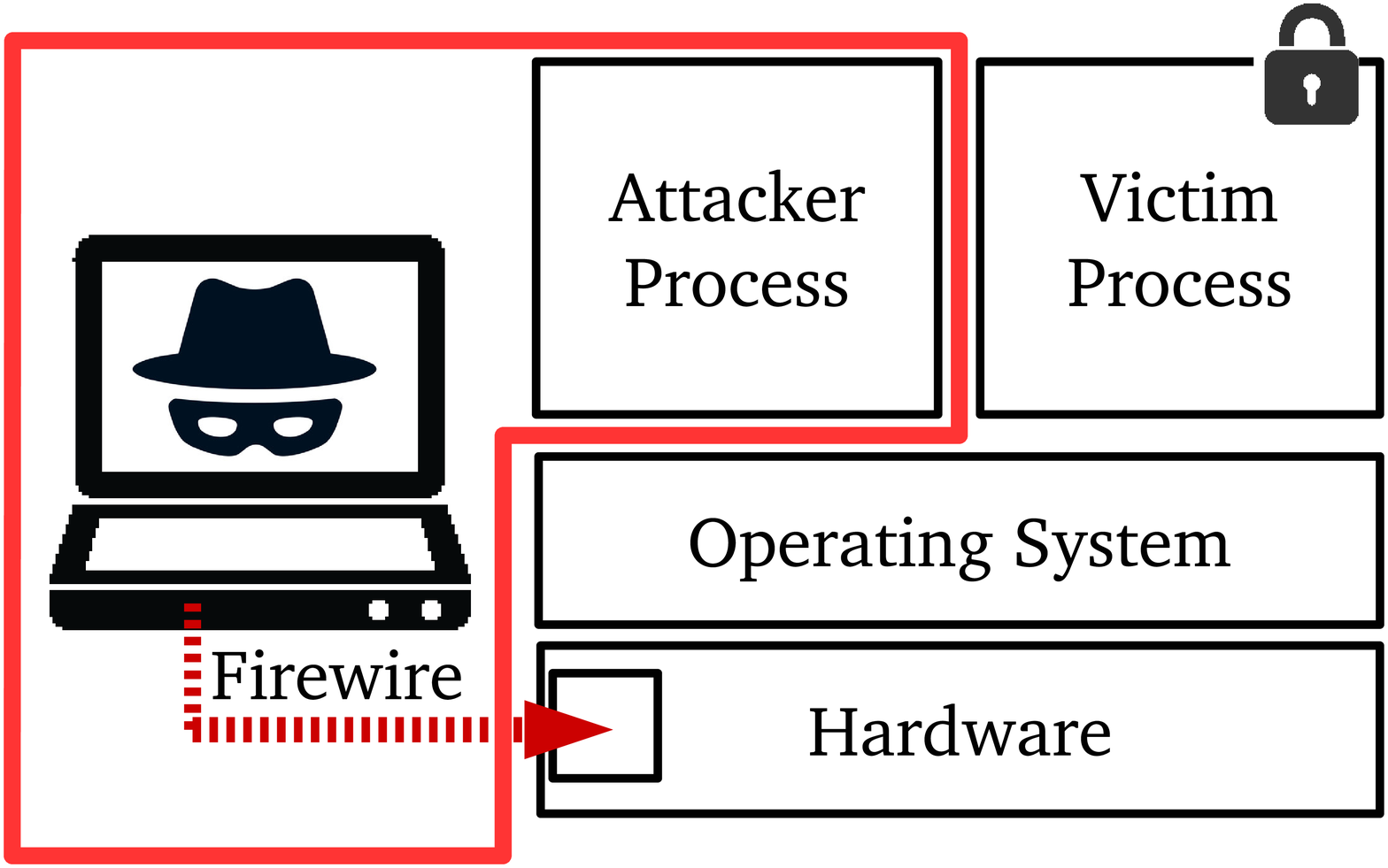}
  \caption{The considered attack scenarios. The red box denotes the capabilities of the adversary.}
  \label{fig:attack}
\end{figure}

Figure~\ref{fig:attack} illustrates the considered attack scenario.
The red box denotes the access boundary of the adversary.
We assume that the adversary was able to install an unprivileged malware process on the system, but has no root privileges on it.
The red arrow illustrates that the adversary can also physically access the platform (e.g. plug in a USB stick or connect a firewire device).
Of course, we suppose that the victim is aware of the valuable assets on his compute platform,
and has therefore activated main memory encryption to protect specific processes.
It is important to note that we do not assume any vulnerabilities in the underlying OS kernel.
We also do not assume that the adversary and the victim necessarily share CPU cores.

In particular, this leads to the following assumptions:
a) the memory is encrypted, i.e., the adversary does not know what values are encrypted;
b) the memory accessing tools can retrieve only ciphertext, and the adversary has no access to the encryption keys;
c) the adversary has the ability to modify memory locations using a physical device (as described in Section \ref{sec:introduction}).
Since he modifies only ciphertext, the modifications lead to some kind of unpredictable corruption of the plaintext. 

\section{Software Based Main Memory Encryption}
\label{sec:soft_seal}
We implemented a software based main memory encryption scheme to replicate the functionality of AMD's SME (due to the yet unavailability of these hardware extensions).
It works transparently without any need to modify the running applications.
We used AES as the block cipher and leveraged the performance speedup that the processor hardware offers via the dedicated AES instructions (AES-NI~\cite{intelaes2012}).

\subsection{Implementation}
\begin{figure}[ht]
  \centering
  \includegraphics[width=.8\linewidth]{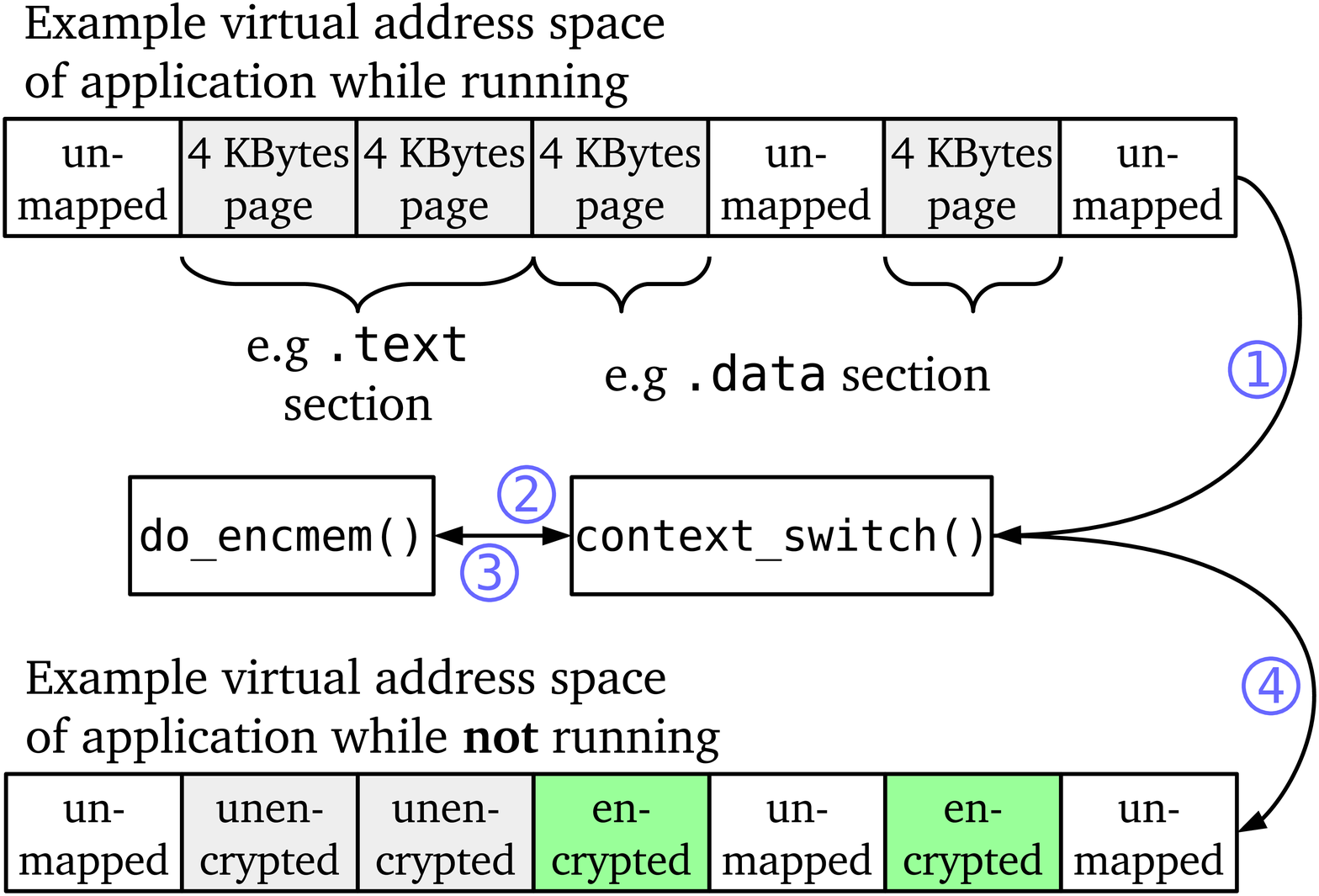}
  \caption{Main memory transparent encryption scheme. When the Linux kernel schedules a new process all present pages of our process are encrypted in place.}
  \label{fig:memory_encryption}
\end{figure}

We wrote a kernel module that notifies the kernel on ``protection worthy'' applications, and extended the Linux kernel itself to perform the encryption.
From the driver, we are able to enable/disable the encryption in general, and to notify the kernel about processes (identified by their process IDs) deserving encryption.
The driver holds a protection list with PIDs, for which the memory pages should be encrypted when the process is currently not running.
Figure~\ref{fig:memory_encryption} illustrates the main encryption procedure.
When the Linux OS decides to schedule a new process, it calls the function \texttt{schedule()}.
In this function, among other things, the OS switches to the new process' memory context ({\tt context\_switch()}~\circled{1}).
We installed a hook in this function, to determine whether the last scheduled process belonged to our protection list. 
If so, we call \texttt{do\_encmem} \circled{2} to encrypt all present (writable) pages of this process.
We limit our encryption to only writable pages because protection worthy material (e.g., keys) is stored in a writable section.
To do this, we need to figure out which pages are actually present in the page table.
The Linux kernel provides a data structure called \texttt{struct mm\_walk}, and several callbacks can be specified to walk the page table within this data structure. 
The \texttt{pte\_entry} function is called for each non-empty PTE (4th-level) entry.
The function \texttt{walk\_page\_range} can then be called to walk over a page range of a specific memory context.
In our case, we walk over the entire 3~GBytes address space of the process%
\footnote{By default Linux' address spaces are split between kernel (1~GBytes) and user (3~GBytes).}.
The callback function is invoked for every valid page entry.
In this callback function, we encrypt each encountered 4~KBytes page in place. 
After all the pages are encrypted, we return to the {\tt context\_switch()} method \circled{3}.
Subsequently, normal execution is resumed, and the memory contents of our process have been encrypted \circled{4}.
We also check whether the next process is in our protection list, and then decrypt all of its writable pages to make it ready for execution.

For simplicity, the encryption code and the necessary keys are stored in RAM, but it does not matter for our demonstration that only tests a fault injection.
Other publications~\cite{muller2011, colp2015} have already shown how to store cryptographic material outside of RAM and also perform cryptographic computation without leaking sensitive material to RAM.
Our scheme could easily adopt such mechanisms.
Still, it is important to note that our implementation does not provide a complete and secure solution by any means.
Its sole purpose is to behave like a hardware scheme and provide a means to demonstrate our fault injection.

\subsection{Software Implementation vs. AMD SME}
Since AMD's processor extensions SME are not available on the market yet, we implemented the software based memory encryption as close as possible to the information AMD revealed thus far~\cite{sme2016, amd2016}.
The requirements for the fault attack to work can be broken down into four properties.
\begin{table}
\centering
\begin{tabular}{|c|l|c|c|}
\hline
\multicolumn{2}{|l|}{\multirow{2}{*}{{\bf Property}}} & {\bf Software} & {\bf AMD} \\
\multicolumn{2}{|l|}{}                                & {\bf impl.}    & {\bf SME} \\
\hline
{\bf 1} & Unit of encryption & 4096~Bytes & 64~Bytes \\
\hline
\multirow{2}{*}{{\bf 2}} & DMA access to enc. & \multirow{2}{*}{yes} & \multirow{2}{*}{yes} \\
  & memory possible & & \\
\hline
{\bf 3} & Memory authentication & no & no (?) \\
\hline
\multirow{2}{*}{{\bf 4}} & \multirow{2}{*}{Encryption enforced by} & Operating & Memory \\
  & & system & controller \\
\hline
\end{tabular}
\caption{Difference between AMD SME and our software prototype.}
\label{table:req}
\end{table}
In Table~\ref{table:req} we show what these properties are and to what extent our software implementation differs with respect to AMD SME.
We now discuss whether or not this impacts the fault attack (with the indexing we refer to the properties as depicted in Table~\ref{table:req}):
\begin{itemize}
\item[{\bf 1}]{
The {\it unit of encryption} determines the required precision of a fault attack and the size of the affected area.
On decryption every bit in the affected plaintext block might be faulty.
Whether this is an advantage or disadvantage for an attacker depends on the situation:
sometimes the target of the injection may lie close to other vital data which an attacker
would like to keep unscathed; at other times locating the target may be more difficult,
so a reduced precision requirement would be beneficial.\\
The attack used in our paper is not affected by the block size, as we can reliably
determine the location of the target prime with sufficient granularity. Furthermore
it is irrelevant for the Boneh-DeMillo-Lipton fault attack how many bits are changed.
}
\item[{\bf 2}]{
According to AMD's documentation {\it DMA} read/write {\it access to encrypted memory} is {\it possible}.
This is of course a core requirement for the attack to work.
}
\item[{\bf 3}]{
All documentation from AMD show that {\it memory authentication} is not applied, because authenticating the entire encrypted memory would cause a substantial storage overhead.
}
\item[{\bf 4}]{
For our software implementation the {\it encryption} is {\it enforced by} the operating system, therefore a determined adversary with kernel-level access could disable the encryption.
However, in this paper we only consider fault attacks on the encrypted memory itself, as our kernel-level
implementation only serves as a vehicle to demonstrate unauthenticated memory encryption. We do not claim
that our pure software implementation provides an equivalent security level to a hardware implementation
inside the memory controller like AMD SME.
Thus, this difference does not impact the attack mechanism.
}
\end{itemize}

\section{Attacking GnuPG}
\label{sec:attack_gnupg}
Our attack combines the fault injection principle with traffic analysis based on cache side channels. It introduces different ways to leverage this combination in order to attack cryptographic applications on a general purpose platform.
To have a concise description, we kept our attack simple and describe only a proof-of-concept.
We solve the practical problems in Section~\ref{sec:real_world_issues}.
To test our attack on a real application, we performed our fault attack on the RSA signing operation in GnuPG Version 1.
As almost all commonly used email clients provide a way to integrate GnuPG into the application to sign and encrypt emails, we deemed this a reasonable target.\\

\begin{figure}[ht]
  \centering
  \includegraphics[width=1.\linewidth]{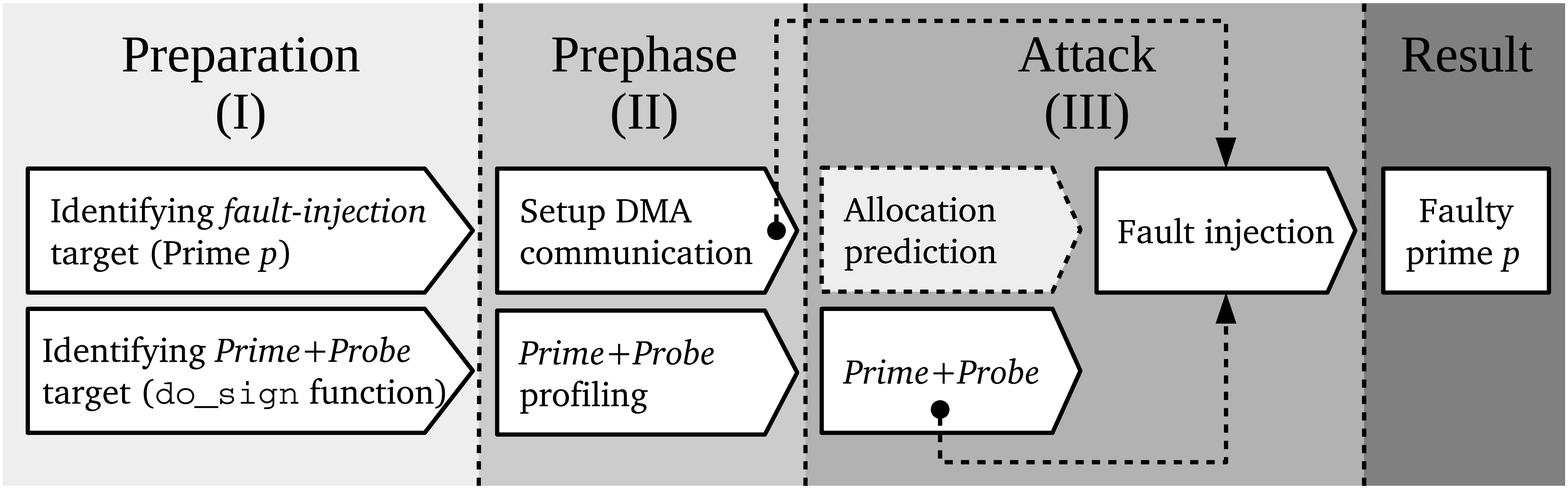}
  \caption{The three phases of our attack on GnuPG.}
  \label{fig:attack_roadmap}
\end{figure}

We first provide a short roadmap on the different components of the attack, because it integrates multiple mechanisms which have to be timed correctly in order for the attack to work.
Figure~\ref{fig:attack_roadmap} gives an overview of each step of the attack.
The details for every step of the attack are given in the following sections.

\paragraph{Preparation: fault injection target}
From Figure~\ref{fig:attack_roadmap}-I can be obtained that the first step is to identify a potential fault injection target.
GnuPG uses the CRT to speed up the exponentiation in RSA signatures (see Section~\ref{sec:pre_not}).
We run GnuPG 1.4.19 with the signature checking disabled\footnote{
The parameter {\tt --no-sig-create-check} disables the signature checking.}.
An RSA key in GnuPG consists of six elements: $n$, $e$, $d$, $p$, $q$ and $q_{inv}$, which are stored in a key file on disk.
So for using the Boneh-DeMillo-Lipton we want to inject the fault into one of the primes {\it p} or {\it q}.
The key file itself is protected with a passphrase (in this case, using 3DES). 
Before the signing operation commences, the user has to type in his password in order to unlock the key.
Only then, the key elements are decrypted to main memory.
Here we stress that the attack window is still wide, because after the key was decrypted from disk a number of computation-heavy operations have to be performed to create the signature.
Among them the exponentiations of $s_{p}$ and $s_{q}$, which have a runtime of $\mathit{\varOmega}(({|n|}/{2})^{2})$.

It is further worth mentioning that our attack can also be modified to apply to DSA and ECDSA operations with some effort.
The attack then targets the respective nonce, aiming to blindly produce a certain bit pattern in its decrypted value. I.e., we follow a well-known lattice-based recovery algorithm from~\cite{naccache2005} to determine the secret key.
Although being more elaborated, this attack has the big advantage of using GnuPG in its (DSA and ECDSA) default mode without utilizing any ``non-default'' GnuPG parameters.

\paragraph{Preparation: Prime+Probe target}
In general \textit{Prime+Probe}~\cite{liu2015} monitors cache eviction.
The adversary selects a number of addresses that fall into the same cache set as the one from the victim binary.
In the \textit{prime} phase the adversary fills the cache sets with his own garbage data.
The adversary \textit{idles} for a few cycles, and then \textit{probes} in the last step.
There, he measures the access time to the addresses that fall into that same cache set as the one from the victim binary.
If the victim process executed at the specific address, it would have evicted some of the adversary's cache lines.
The adversary could observe this via an increased memory access latency for those lines.

So for our attack we need to identify a feasible {\it Prime+Probe} target (Figure~\ref{fig:attack_roadmap}-I).
By inspecting the GnuPG binary we can determine an address in the executable section of the binary where the key is already present in main memory.
In our case, this is the function {\tt do\_sign} in the file {\tt g10/sign.c}.
The virtual address of this function can be determined by inspecting the binary (e.g. by using {\tt objdump}).
By mapping the binary into our address space and again leveraging the {\tt pagemap} we can determine the physical address of that function.
Then, we have to determine where this physical address is stored in the LLC.
Based on the hash function (Section \ref{sec:background}), and the known set bits we can determine what cache slice and set the physical address is assigned to.
Now, we have to determine a number of memory locations that, in terms of cache set and slice collide with the physical address of the {\tt do\_sign} function.
On our test system we have a 6~MBytes twelve-associative LLC, thus we are looking for twelve colliding addresses.
With this information and the knowledge of the hash function (Section~\ref{sec:background}), we can brute-force the calculation over an address window of 6~MBytes to determine twelve addresses that are assigned to the same cache slice and set.

\paragraph{Prephase: setup DMA communication}
In the prephase of the attack we need to setup the DMA communication (Figure~\ref{fig:attack_roadmap}-II)
As described in Section~\ref{sec:attack_model} we defined that the adversary was able to spawn a process on the same host, but has also connected a remote DMA device (e.g. laptop).
Now, in order to inject the fault at the right time the adversarial process has to notify the external DMA device about when to inject the fault.
To do this, the adversarial process allocates a piece of memory and determines the physical address of the allocation using the {\tt pagemap} (details are deferred to Section~\ref{sec:real_world_issues}).
The determined location is then sent to the external agent (who can then set his DMA device to this address location).
Once negotiated, the adversarial process uses this memory location to notify the external DMA device when to inject the fault (by writing a specific pattern to this location)\footnote{
When using a DMA device to inject the fault, the RAM access is still performed by the memory controller through the root complex~\cite{funderbolt2013}, therefore ECC (Error Correcting Code) is irrelevant.}.
The adversary can of course also notify the external agent via network, but depending on the configuration of the system and the requirement for stealthiness of the attack this might be problematic.

\paragraph{Attack: Prime+Probe}
To successfully carry out the {\it Prime+Probe} attack we have to calculate the mean access time over all twelve addresses.
If this exceeds a certain threshold, we know that one of the addresses has been evicted from the cache, probably because the victim process has reached the {\it do\_sign} function in its execution.
In order to achieve this, we execute the algorithm shown in Figure~\ref{algo:prime_probe} in a tight loop.
The threshold values are determined experimentally in a prephase of the attack (Figure~\ref{fig:attack_roadmap}-II), greater details on exact values for our attacked platform are given in Section~\ref{sec:attack_results}.
\begin{figure}
\begin{algorithmic}
\WHILE {1}
  \STATE delta\_mean = 0\;
  \FOR{$i = 0$ \TO cache\_ways}
    \STATE delta[i] = probe(addr\_collision[i])\;
    \STATE delta\_mean += delta[i]\;
  \ENDFOR
  \STATE delta\_mean = delta\_mean/cache\_ways;
  \IF{ delta\_mean $>$ 140 \AND delta\_mean $<$ 180}
    \IF {pause $\geq$ 150000} 
      \STATE return 1\;
    \ENDIF
    \STATE pause = 0\;
    \STATE break\;
  \ELSE
    \STATE pause++\;
  \ENDIF
\ENDWHILE
\end{algorithmic}
\caption{Algorithm to perform the {\it Prime+Probe} attack}
\label{algo:prime_probe}
\end{figure}
The variable {\it delta\_mean} holds the mean access time over all twelve memory accesses.
We added the {\it pause} variable to prevent false positives, and since the loop runs without any delays at full CPU speed we had to set the variable's threshold value quite high (150000).
However, this value was determined manually and varies greatly depending on the target platform.
We had to run a large number of experiments.
For every iteration we determined the number of detected cache accesses, if the number was greater then one, we adjusted the value accordingly.

\paragraph{Attack: fault injection}
The actual attack starts by checking whether GnuPG was started, and then beginning to do the {\it Prime+Probe} (Figure~\ref{fig:attack_roadmap}-III).
When the attack process determines that GnuPG executes the {\tt do\_sign} function, we enforce a schedule call.
GnuPG is then put in to the background, and its memory gets encrypted.
This is of course just a vehicle because our software-based main memory encryption only works for processes in the background.
In a real hardware implementation this would not be necessary.
We then have to determine the location where to inject the fault.
To do so we profiled GnuPG beforehand to determine the location of the key structure in memory (Figure~\ref{fig:attack_roadmap}-III ``Allocation prediction'').
We predict the physical location of the key structure using a PFN leakage mechanism similar to the one described in~\cite{kemerlis2014} (again the details are described in Section~\ref{sec:real_world_issues}).
We then use the remote DMA to inject the fault (Figure~\ref{fig:attack_roadmap}-III).
For this purpose we extended the Inception framework~\cite{inception2016} to be able to read and write memory via a FireWire cable (limitations to this approach and countermeasures are discussed in Section~\ref{sec:dma})\footnote{%
Important to note here is that new attacks such as Rowhammer~\cite{seaborn2015} could also be leveraged to manipulate the memory and inject the fault.}.
After the fault has been injected, we resume the execution of GnuPG.
The kernel will decrypt all memory pages of GnuPG to make it runnable again, among them the one with the modified memory location.
Once decrypted the value of $p$ will be faulty.
GnuPG will then create the faulty signature.
Finally, we calculate $q$ offline based on the obtained faulty signature (see Section~\ref{sec:pre_not}).

\section{Real-world challenges}
\label{sec:real_world_issues}
In Section~\ref{sec:attack_gnupg} we described a fault attack against GnuPG. In order to convert this proof of concept into a real world attack, the adversary faces four challenges:
\begin{enumerate}
  \setlength{\itemsep}{-2pt}
  \item{Determine {\it Prime+Probe} target addresses.}
  \item{Physical addresses of dynamic data structures.}
  \item{Obtain two signatures of the same message.}
  \item{Find suitable hardware interfaces.}
\end{enumerate}
We address these challenges in the following sections.

\paragraph{Determine Prime+Probe target addresses}
In general the adversary is looking into ways to obtain the translation of virtual to physical addresses for the use in his {\it Prime+Probe} attack, and also for the fault injection.
Which addresses the adversary wants to obtain, is of course very specific to the attacked target.
In our case, this is the \texttt{do\_sign} function of GnuPG.
Obtaining the virtual address of this function is quite easy (as shown in Section~\ref{sec:attack_gnupg}).
The major Linux distributions only slowly adopt position independent binaries.
For now we can safely rely on the virtual addresses we can obtain when inspecting the binary with, e.g. {\tt objdump}.

The {\tt /proc/<pid>/pagemap} file can be used to obtain a physical address.
For every user-space page, the {\tt pagemap} provides a 64 bit value, indexed by its virtual page number, which contains information regarding the presence of the page
in RAM.
Bit 63 determines whether the page is present in memory and bits 0 to 54 encode its page frame number.
Under the assumption that the underlying memory is shared between adversary and victim, the adversary can just use {\tt mmap} to map the GnuPG binary into his address space.
Then he uses the {\tt pagemap} to determine the physical address of the function.

Other ways to obtain the desired information is to look for specific cache access patterns that relate to the execution of this function.
Trace driven cache attacks have been already discussed (e.g.,~\cite{aciiccmez2006}.
They trace cache-set activity, and look for the access pattern in order to obtain information about the inner state of an encryption algorithm.
In \cite{liu2015}, the authors show how to perform a practical attack based on the LLC.
Based on these publications, the recorded cache access pattern can directly be used to identify that the process is currently executing at a certain address.
Alternatively, other parts of the cache can be monitored at the same time, to identify the actual cache set that a specific address falls into.

\paragraph{Physical addresses of dynamic data structures}
The location of the key data structure (that includes $p$) is not only unknown,
but also differs for every execution, because it is stored on a writable page (on the heap).
To obtain the physical address of this dynamic data structure an adversary can again draw on the {\tt pagemap} mechanism.

The adversary can easily trace a single execution run of GnuPG with the desired commandline parameters
and concurrently monitor the pagemap for new page allocations.
Mere syscall tracing is insufficient, as the adversary needs to know the actual
physical memory footprint of the process, not the number and size of allocations.
As Linux employs a lazy allocation strategy, page ranges which have been requested by
\texttt{mmap} but not touched will not have physical backing at all, whereas other
areas that do not require explicit allocation (e.g. stack growth) may indeed cause
the number of consumed physical pages to increase.
Once the adversary has reached the point where $p$ is copied into memory, the trace is complete.

In our case, we determined that $p$ will be placed on the 10th allocated page.
This knowledge still does not tell the adversary the exact physical address where
the key will be located in a future run of GnuPG.
However he can take advantage of the fact that the Linux kernel allocator
tends to give out pages that where freed shortly before.
So, most recently freed memory pages are given out first to processes.
Thus the adversary can allocate a number of physical memory pages (i.e. allocate
them and actually perform write operations to them), consult the {\tt pagemap}
to determine their physical addresses, and free the pages again.
Since the Linux kernel will most likely give out these very same pages to GnuPG
the adversary can predict which physical page will contain the key.
Freeing the pages has of course to be timed correctly by the adversary so that no other program gets these pages.
But this can easily be combined with the already running {\it Prime+Probe} attack.

Our observations also showed that if our prediction fails, i.e. the 10th page
allocated to GnuPG did not match the 10th from last page freed beforehand, that
no newly allocated page matched any freed page at all.
Also the accuracy of the prediction astonishingly depends on the overall number
of pages the adversary allocated before. Both of these findings are worth
futher investigation.
The exact success probability of such an attack is presented in Section~\ref{sec:attack_results}.

\paragraph{Obtaining two signatures of the same message}
There are several ways an adversary could obtain a signature for the same message twice.
But first there is an obstacle the adversary has to overcome.
Along with the message that should be signed GnuPG puts a Unix timestamp into the hash calculation.
Fortunately for the adversary, the timestamp is only in second granularity.
So, when the adversary is able to motivate the victim to create two signatures from the same message shortly one after the other he gets the desired double signature.

There are a number of scenarios the adversary could leverage.
Any form of signed \textit{auto-reply} message (e.g.~\textit{out-of-office} notifications) would work for the adversary.
It is possible to configure all commonly used email clients (e.g. Outlook, Thunderbird, Apple Mail, etc.) to sign \textit{auto-reply} messages with a defined key.
The GnuPG options \texttt{--passphrase-file}, \texttt{--passphrase-fd} or \texttt{--passphrase} allow a user to provide the passphrase for the signing key either directly from the commandline, some file or from a file descriptor.
It is questionable, from a security perspective, to store the passphrase for the key in a file, however, such scenarios exist.

Moreover, for both cases services such as GoodCrypto~\cite{goodcrypto2016} or CipherMail~\cite{ciphermail2016} provide transparent email encryption directly on the email gateway (independent from the email clients used by the communicating parties).
This obviates the need for non-tech-savvy users to configure PGP individually on their end systems.
Instead, the keys are stored on the server, and emails are transparently encrypted by the email gateway with the appropriate key based on the sender's email address.
As these encryption plugins are stateless and do not inspect the mails (e.\,g., to skip automatically generated emails), adversaries can again provoke repeated signatures of the same message.

\paragraph{Find suitable hardware interfaces}
In our attack we rely on the FireWire interface.
One could argue that modern systems might not be equipped with a FireWire controller anymore.
However, our attack can not only be performed through a native FireWire port, but also via ExpressCard/PCMCIA expansion ports or a Thunderbolt to FireWire adapter.
It is likely that a system has at least one of the aforementioned interfaces.

\section{Attack results}
\label{sec:attack_results}
All experiments were conducted on a Lenovo Thinkpad T520 with an Intel Core i7-2670QM CPU and 4~GBytes of RAM.
This platform represents a standard off-the-shelf laptop which uses the aforementioned hash function for LLC slice determination.
The device is equipped with a FireWire interface, and we verified the above described memory modification capabilities.
Software-wise we used Debian 8 ``Jessie'' with a Linux kernel version 4.0.

\subsection{Prime+Probe results}
For our attack we had to determine a threshold for the cache eviction in order to perform the \textit{Prime+Probe} attack.
All accesses were measured using the \texttt{rdtscp} instruction\footnote{
\texttt{rdtscp} is a serializing call that prevents reordering around the call, and returns the number of executed processor cycles.}.
We divided our measurements into two sets $\left\{ A \right\}$ and $\left\{ B \right\}$.
\begin{itemize}
  \setlength{\itemsep}{-2pt}
  \item Set $\left\{ A \right\}$ was our baseline: we chose twelve set-colliding memory locations, accessed all of them to allocate them into the cache, and then immediately measured the access time for all twelve locations again.
  \item For set $\left\{ B \right\}$, we chose twelve memory locations that fell into the same cache set and slice as the address we want to monitor.
  We accessed all of them once, then we executed one instruction from our victim application (GnuPG) at exactly that address, and finally measured the access time for our twelve lines again.
\end{itemize}

\begin{figure}
  \includegraphics[width=1.\linewidth]{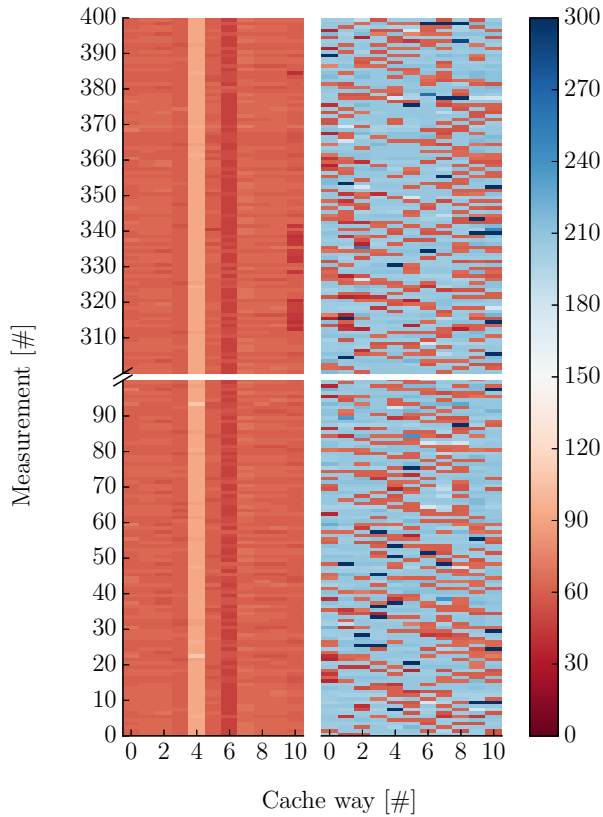}
  \caption{{Cache way access times measured with {\tt rdtscp}.}}
  \label{fig:cache_access1}
\end{figure}

\paragraph{Cache ways}
First, we had to figure out if we are able to reliably measure variations in access times to addresses from the same cache set.
Figure~\ref{fig:cache_access1} illustrates this experiment.

Set $\left\{ A \right\}$ is represented by the left column.
Our assumption was that the measured access time would be very low, because all twelve locations fit into the cache. Since no other application ran in between, all access requests would be served by the cache.
In \cite{levinthal2009}, Levinthal shows that when an access request to an address takes less than 40~cycles, it indicates that it was served from the cache: $\sim$4~cycles indicate that the request was served by the L1 cache, $\sim$10~cycles indicate an L2 cache access, and $\sim$40~cycles indicate a load from the LLC.
Indeed, all access times were in the range between 0 and 40~cycles.

\begin{figure}
  \includegraphics[width=1.\linewidth]{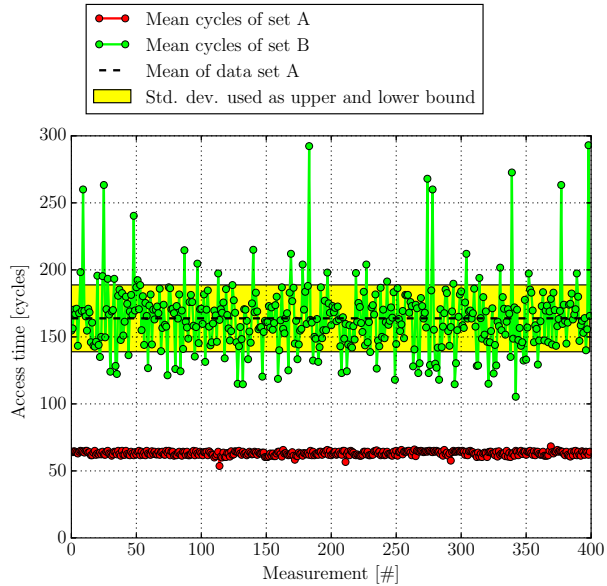}
  \caption{The mean access time over 400 measurements.}
  \label{fig:mean_access}
\end{figure}

Set $\left\{ B \right\}$ is represented by the right column.
Our assumption was that at least one of the addresses had to be evicted from the cache, because the cache only has twelve ways.
Indeed, Figure~\ref{fig:cache_access1} supports this assumption.
The processor had to load some addresses from main memory.
We assume that more than one address was evicted from the cache, because our measurements were conducted from a separate application, so not only the victim code ran in between but also other kernel scheduling code, etc.
Also cache prefetching behaviour might played a role here.
This explains why more than one address was fetched from the main memory (light blue spots in the figure).
However, in general we confirmed that we would be able to measure when the victim application executed its binary at a certain memory location.

\paragraph{LLC access}
So far, we confirmed that we could recognize cache eviction in terms of cache ways.
To know when a certain application hit a certain address in the executable, we had to specify a reliable threshold.
To this end, we calculated the mean value over all twelve access times for set $\left\{ A \right\}$ and for set $\left\{ B \right\}$.
Figure~\ref{fig:mean_access} illustrates the result, each showing the mean access times over all twelve addresses.
Clearly, the mean access time is significantly higher when one of the addresses was accessed.
Therefore, we were able to set the threshold to $delta\_mean = mean(\left\{ A \right\}) \pm std(\left\{ A \right\})$ (see Figure~\ref{algo:prime_probe}), corresponding to the yellow bar in Figure~\ref{fig:mean_access}.
The noise we see in the figure is due to the fact that the victim's data can be partially cached in higher-level caches (leading to faster reads), and the variation in the L1 and L2 contents affects the cache probe time and induces measurement noise.

\subsection{Page allocator prediction results}
As already described in Section~\ref{sec:real_world_issues} it is necessary to find the physical address of the prime factor {\it p}.
As it is allocated on the heap it is necessary to somehow predict the right physical address where {\it p} is located on.
To predict the correct physical address we did the following experiment.
First we annotated GnuPG to print the virtual and physical address of the prime {\it p}.
Then, in our adversarial process, we allocated a number of pages using {\tt mmap} and calculated their respective physical address (using the {\tt pagemap}).
Afterwards we freed all these pages and let GnuPG run.
We then compared if the physical address of the prime {\it p} was among our previously allocated and freed pages.
We did this with a various number of allocations, ranging from only 8 allocations up to 1024.
\begin{figure*}
  \centering
  \includegraphics[width=1.\linewidth]{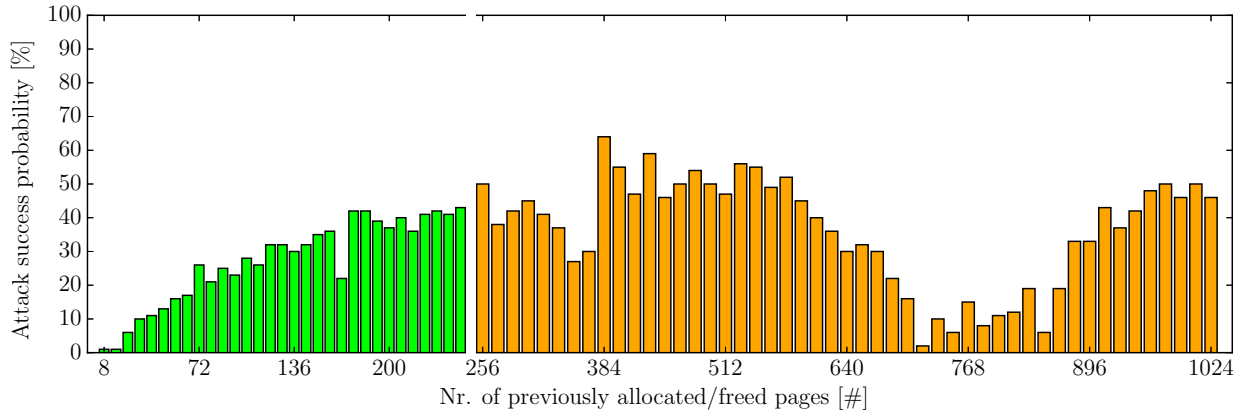}
  \caption{Attack success probability based on the previously allocated and freed memory pages.}
  \label{fig:heap_predict}
\end{figure*}
The result can be obtained from Figure~\ref{fig:heap_predict}.
The figure is split into two parts. The first green bars contain results in steps of 8, after 248 experiments we increased the number of allocations in every measurement run by 16.
So the orange bars contain the results from 256 to 1024 in steps of 16.
For every allocation step we performed 100 measurements.

What we observed is quite interesting.
When the physical address of the prime was among the previously allocated/freed pages it was always on the same one.
In our case it was always the tenth last page from our pool of allocated/freed pages.
So it was entirely deterministic which of the previously allocated/freed pages would later contain the prime {\it p}.
But as can be obtained from the figure only in a certain number of measurements the physical page of the prime was among the allocated/freed pages at all.
Moreover, the overall success rate depended on the number of previously allocated/freed pages.
We achieved a maximum success rate of $\sim$55-60\% when allocating and freeing between 380 and 500 pages
before executing GnuPG.
This value is already surprisingly high, given that the window of uncertainty (the time between the release
of the pages and GnuPG receiving a page for $p$) includes the creation of a process and thus a full address space.
A determined attacker could employ the techniques used in this paper to release pages much closer
to the critical allocation, thus further boosting his chances.

Of course when performing the actual fault injection GnuPG will not happily print the address of the prime {\it p}.
The adversary will blindly inject the fault into a potentially wrong memory
location which can have unpredictable effects on the system behaviour.
In our case the GnuPG application sometimes just crashed.

\section{Countermeasures and Related Work}
\label{sec:counter_related_work}
In the following section we present a number of topics which are relevant for this work.
We also discuss countermeasures for each attack vector.

\subsection{Cache-based side channels}
\label{sec:cache_side_channel}
Cache-based side channels have a long history~\cite{percival2005, tiri2007, osvik2006}.
In recent years, the focus has been shifted toward the LLC~\cite{yarom2014, gruss2015, maurice2015, hund2013}, which is typically shared between multiple cores, leading to new forms of side channels and attacks.
But these attacks are harder to carry out, because, e.g., Intel uses an undocumented hash function to distribute addresses to different segments of the LLC.
Thus, initially, a lot of effort has been put into reverse engineering this hash function for different processor generations~\cite{maurice2015, hund2013}.
Based on these new findings, a number of attacks have been proposed.
In~\cite{liu2015} Liu et. al show how to snoop on processes over VM boundaries.
Yarom et al.~\cite{yarom2014} show how to extract the private encryption keys from a victim program in a single operating system and between processes running in separate VMs.\\

Mitigating cache-based side channels remains a challenging task.
In~\cite{kong2013}, Kong et al. investigate several schemes to mitigate cache attacks.
Also, modern Intel server CPUs provide a technology called CAT (Cache Allocation Technology)~\cite{intel2015} to prevent traffic analysis by other processes or VMs through eavesdropping the LLC.
CAT allows an OS or a VMM to control the allocation of the shared LLC.
Once CAT is configured, the processor allows access to portions of the cache according to the established class of service.
The processor obeys to these classes when it runs a process.

\subsection{procfs}
The procfs provides user applications with useful information about the current system state.
However, often these information can be used as a means to spy on other processes or resources in a system.
Jana and Shmatikov~\cite{jana2012} show how to exploit information provided by procfs to create detailed profiles of applications.
They use memory consumption discrepancies of a browser to determine which website the user currently looks at.
Zhou et al.~\cite{zhou2013} leverage, among other sources, the procfs and sysfs to gather detailed information about the user of an Android smartphone.
Along with the usual nodes in procfs Android introduces new ones that reveal even more information about the system or user (e.g. data usage, location, etc.).\\

procfs is an important resources in Linux' system architecture.
Many applications (e.g. {\tt ps}, {\tt netstat}) rely on information exported through procfs.
Therefore disabling this filesystem is not an option.
However, as already done with many other files in procfs the access to some security critical files should be limited to admin users.
For the {\tt pagemap} this already happend.
Since kernel version 4.1 the interface {\tt /proc/self/pagemap} can only be used if the user has the kernel capability {\tt CAP\_SYS\_ADMIN}.
First the Linux kernel developers blocked the access to the file entirely.
A normal user was not even able to open the file.
Since Linux kernel version 4.2 the file can be opened by normal users, but again only if the one reading from the file posses the capability {\tt CAP\_SYS\_ADMIN} can he actually read content from the file.
Otherwise, he will just read zeros.
In user space this basically translates to, that only admin users can read from this interface to prevent the leakage of this security critical information.
The access to other security critical information that are exported through procfs, e.g., {\tt kallsyms} were also limited in the past.

\subsection{DMA}
\label{sec:dma}
In order to launch our fault attack we rely on DMA.
A lot of attacks have been proposed to launch a DMA based attack using various interfaces~\cite{becher2005, wojtczuk2008, dornseif2004}.
In 2006 Boileau~\cite{boileau2006} showed the remote DMA capabilities of the Firewire bus.
More recent attacks have been proposed by Sevinsky~\cite{funderbolt2013} in 2013.
Sevinsky used the Thunderbolt interface to launch a DMA attack.\\

The obvious countermeasure to prevent DMA attacks is using an IOMMU~\cite{iommu2014, amd2015}.
However, when running a 32Bit Linux kernel the IOMMU (if present) never gets enabled.
On 64Bit systems the Linux kernel makes use of the IOMMU to block remote DMA accesses.
However, it is highly processor specific whether the system contains an IOMMU or not.
If no IOMMU is present users are advised to disable the bus master capabilities of specific devices.
Which device can function as a bus master can be determined via {\tt lspci}.
Disabling the bus master capabilities on Linux can be done via the {\tt config} file for the specific device in the sysfs.

\subsection{Fault attacks}
Fault attacks are a well-known concept in computer security.
Initial work on fault attacks was done by Boneh et al.~\cite{boneh1997} in 1997.
In this groundbreaking work they show that various cryptographic algorithms can be attacked using hardware fault injection.
In particular they attack the Fiat-Shamir scheme and Schnorr's identification scheme.
Since then, several fault attacks have been proposed.
In 2003 Dusart et al.~\cite{dusart2003} describe a differential fault analysis attack on AES.
They are able to break an AES128 key with around 10 faulty messages.
In the same year Müller et al.~\cite{aumuller2003} propose an attack on RSA when using CRT.\\

In the early work of Boneh et al.~\cite{boneh1997}, the authors already propose the use of random padding%
\footnote{It is interesting that GnuPG version 1 is not using random padding.}
to protect against their attack (as suggested by Bellare and Rogaway~\cite{bellare1996}).
In 1999 Shamir~\cite{shamir1999} came up with a scheme to protect against fault attacks.
However, Shamir’s scheme only protects the signature computations modulo the two secret prime factors of the RSA modulus $n$.
The CRT combination step to obtain the final signature modulo $n$ is left unprotected.
Indeed, all known mitigation techniques against fault attacks apply here.
An effective countermeasure that is already discussed in \cite{boneh1997, bellare1996} is the addition of cryptographic padding (e.g. OAEP~\cite{bellare1994}) to the message.
Later, effort has been put into designing fault resilient cryptographic algorithms.
In~\cite{giraud2006} Giraud proposes an RSA implementation which is resistant to fault attacks.
Kim et al.~\cite{kim2007} propose a side-channel analysis and fault attacks resistant RSA-CRT implementation.
The novelty of their approach is how to find the best solution to prevent side channel attacks and fault attacks with a minimal efficiency loss in RSA-CRT.

\subsection{Main memory encryption}
Henson et al. \cite{henson2014} provide a survey on memory encryption techniques.
Other work on main memory encryption has been done by Chhabra et al~\cite{chhabra2011}.
They propose an encryption scheme for systems with NVMM (non-volatile main memory).
Their threat model is one in which an adversary obtains physical access to the system and extracts sensitive information from the storage system by reading it.
The attack scenario is valid for mobile devices, where an adversary would be able to easily read out the RAM cells.
However, the scheme would not keep stronger adversaries at bay, which are able to manipulate parts of the encrypted memory.
In \cite{muller2011}, Müller et al. propose an architecture called TRESOR.
TRESOR makes cold boot attacks difficult, because instead of using RAM, it ensures that all encryption states as well as the secret key and any part of it are only stored
in processor registers, thereby substantially increasing its security.
Based on the approach of TRESOR and an encryption scheme very similar to the one we implemented, Götzfried et al.~\cite{gotzfried2016} propose a RAM encryption scheme called RamCrypt.
However, since the scheme uses AES in XEX (Xor-Encrypt-Xor) mode of operation and as an IV the virtual address, the scheme suffers from the same vulnerability as the scheme that we study here, because it does not perform memory authentication.
In 2010, Champagne and Lee~\cite{champagne2010} proposed Bastion, a security architecture whose goal is to protect the execution and storage of trusted software modules within untrusted commodity software stacks.

\subsection{Memory authentication}
Our memory encryption scheme could be extended by memory authentication.
We could hold a hash for every encrypted page in a process.
On decryption the encryption engine would calculate a hash over the actual memory area and compare it against the stored hash.
We have not investigated this approach thoroughly and the additional runtime overhead from the authentication remains to be measured.
Moreover, the amount of additional storage for the hashes also needs to be quantified.\\

An integrity tree is the classical way to provide authentication to a larger amount of data.
Elbaz et al.~\cite{elbaz2009} provide an overview of several concepts for memory authentication.
Among them are classical concepts like Merkle Trees.
These are binary trees where each node holds a hash digest of its two children, and the lowest leaves hold digests of the protected data units.
Our memory encryption scheme could be extended to hold such a tree.
In 2007, Rogers et al.~\cite{rogers2007} propose Bonsai Merkle Trees a new organization of the Merkle Tree that reduce its memory storage and performance overheads.
The authors manage to reduce the performance overhead of the memory integrity verification from 12.1\% to 1.8\% across a number of benchmarks.
Also the amount of storage overhead was reduced 33.5\% down to 21.5\%.
Still, also a memory overhead of $\sim$20\% might be infeasible in many use cases.

\subsection{Signature verification}
It is important to note that we did not exploit any vulnerability in GnuPG.
In default operation mode, GnuPG verifies the generated signatures (see Section~\ref{sec:pre_not}), and if an error is detected, it terminates with an error report without releasing the (faulty) signature.
However, the command line parameter \texttt{--no-sig-create-check} exists, and made our attack possible.
Fortunately, this parameter exists only in GnuPG version 1, and GnuPG version 2 removed the command line option completely.
The OpenSSL library always verifies a signature before releasing it, and repeats the computation without using CRT in case an error is detected.
There is no option to disable the check other than by modifying the source code, and this type of threat is not part of our attack model.
Since the overhead of the signature verification is very small compared to the overhead of the signing procedure, the cost of this mitigation is negligible.
We believe any application or library should implement this check.

\section{Conclusion}
\label{sec:conclusion}
Threat scenarios that include an active adversary on a dynamic system require memory encryption for privacy, and memory authentication for integrity.
The scenario became even more relevant due to the fact that hardware vendors now acknowledged this threat and therefore provide solutions.
So the question addressed in this paper is the following.
Is it reasonable (and to what extent), in order to save the high cost of dedicated authentication mechanisms, to rely on encryption to protect both privacy and integrity?

To answer this question, we showed the possibility of fault attacks on memory that is encrypted with no authentication.
Of course, our attack is complex, and implementing it even as a proof-of-concept was a serious challenge.
The complexity results from the protection that the encryption provides by itself.

Nevertheless, this work clearly illustrates that sophisticated attacks against the integrity of the memory cannot be dismissed, or at least cannot be ruled out.
We therefore propose that memory encryption techniques should include integrity protection, despite the added complexity and performance costs.


\bibliographystyle{abbrv}
\bibliography{paper}

\end{document}